\documentclass[%
 aip,
 jmp,%
 amsmath,amssymb,
 reprint,%
]{revtex4-2}
\usepackage{graphicx}
\usepackage{dcolumn}
\usepackage{bm}
\usepackage{multirow}
\usepackage{xr-hyper}
\begin{document}


\title[W2V to Transformers]{From Word2Vec to Transformers: Text-Derived Composition Embeddings for Filtering Combinatorial Electrocatalysts}

\author{Lei Zhang}
 \email{lei.zhang-w2i@rub.de}
\author{Markus Stricker}%

\affiliation{ 
Interdisciplinary Centre for Advanced Materials Simulation, Ruhr University Bochum, Universitätsstraße 150, 44801 Bochum, Germany
}%

\date{\today}

\begin{abstract}
Compositionally complex solid solution electrocatalysts span vast
composition spaces, and even one materials system can contain
more candidate compositions than can be measured exhaustively. Here we
evaluate a label-free screening strategy that represents each
composition using embeddings derived from scientific texts and
prioritizes candidates based on similarity to two property concepts. We
compare a corpus-trained Word2Vec baseline with transformer-based
embeddings, where compositions are encoded either by linear
element-wise mixing or by short composition prompts. Similarities to
`concept directions', the terms conductivity and
dielectric, define a 2-dimensional descriptor space, and a
symmetric Pareto-front selection is used to filter candidate subsets
without using electrochemical labels. Performance is assessed on 15 materials
libraries including noble metal alloys and multicomponent oxides. In
this setting, the lightweight Word2Vec baseline, which uses a
simple linear combination of element embeddings, often achieves the
highest number of reductions of possible candidate compositions while staying close to the best measured
performance.
\end{abstract}

\keywords{word embeddings, transformers, text mining, composition representation}
\maketitle


\section{Introduction}\label{sec:intro}
Compositionally complex solid solution materials, including high-entropy alloys and multicomponent oxides, provide a large compositional design space for electrocatalyst discovery because small changes in elemental identity and ratio can shift electronic structure, surface chemistry, and charge transport~\cite{Löffler202126894}. Combinatorial synthesis and high-throughput electrochemical measurements enable comparatively rapid exploration of this space~\cite{Ludwig2019Discovery}, but even a single library can contain hundreds to thousands of distinct compositions prepared under specific processing conditions~\cite{Yao20206316,Baeck2002563}. As library size and chemical diversity increase, deciding which compositions merit further measurement or deeper characterisation becomes a central practical step in experimental workflows.

Screening strategies in this setting must balance two needs that pull in different directions: reducing the number of candidates enough to save experimental effort while still keeping at least one near-top performer for a given material system and operating condition. Supervised learning can be effective when labelled data are abundant, consistent, and comparable across libraries~\cite{Pandit20227583,Liu2021}, yet in practice measurements are often sparse, material system-specific, and influenced by preparation details. This makes it attractive to consider approaches that can operate with weak or no labels and that reuse information already present in the scientific record.

Scientific texts provide a compact representation of community knowledge linking composition to property language that is not a direct proxy for electrochemical response but often reflects underlying physical constraints and correlations~\cite{Tshitoyan201995,Mahbub2020}. The materials and electrocatalysis literature repeatedly connects composition to descriptors such as electrical conductivity, dielectric behavior, oxidation tendencies, and other recurring concepts~\cite{Sun20247392,Chen201711969,Qiao201612658}. If compositions can be embedded into a latent space derived from text, then similarity to selected concept directions can define a low-dimensional descriptor space for ranking and filtering candidates in a way that is transferable across different libraries~\cite{zhang2025electrocatalyst}.

At the same time, a text-guided filter depends sensitively on how a multicomponent composition is represented within the embedding model~\cite{Qu2024,Lei20241257}. One option is to represent a composition as a concentration-weighted mixture of element embeddings, which is computationally straightforward and naturally supports fast screening, but it assumes linearity and independence across elements~\cite{Park2024601}. Another option is to embed a formatted full-composition prompt as a single piece of text, allowing the encoder to represent stoichiometric ratios and multi-element motifs more directly, but introducing choices about prompt design and model behavior~\cite{Liu2024240}. Different embedding backbones further impose different inductive biases: lightweight distributional models trained on a focused corpus emphasise co-occurrence structure~\cite{levy_goldberg_2014_sgns_pmi}, while transformer encoders may inject broader contextual associations learned during pretraining~\cite{devlin-etal-2019-bert}. These modelling decisions shape both how aggressively candidate compositions in a given materials space, e.g. a materials library, can be reduced and how reliable high-performing compositions remain in the retained subset that comprises `the prediction'.

In this work, we construct text-derived composition embeddings using Word2Vec~\cite{Mikolov2013,mikolov2013efficient,Goldberg2014}, MatSciBERT~\cite{Gupta2022}, and Qwen~\cite{qwen} in both element-wise and full-composition prompt forms, project each composition onto concept directions related to electrical conductivity and dielectric behavior, and apply a dual Pareto-front filter to select candidates. We evaluate the resulting trade-offs between best-current retention and candidate reduction across 15 combinatorial libraries spanning the hydrogen evolution reaction (HER), oxygen reduction reaction (ORR), and oxygen evolution reaction (OER).

\section{Methods}\label{sec3}

\subsection{Corpus collection (Word2Vec baseline)}

For the Word2Vec baseline, scientific abstracts related to electrocatalysis, high-entropy alloys and complex oxides are collected using the \texttt{PaperCollector} module of the \texttt{MatNexus} framework~\cite{Zhang2024}. Abstracts are sourced from Scopus and arXiv, restricted to open-access publications available up to the year 2024. Only English-language abstracts are included. Metadata and full-text abstracts are automatically retrieved and stored in structured CSV format. This corpus is used exclusively to train the Word2Vec model; the transformer-based models (MatSciBERT and Qwen variants) are used in inference-only mode without additional training or fine-tuning on this corpus.

\subsection{Text preprocessing (Word2Vec baseline)}

Raw abstracts are processed using the \texttt{TextProcessor} module in \texttt{MatNexus}. Preprocessing includes removal of licensing statements and publisher-specific fragments (for example, ``©'' strings or publisher boilerplate), standard stopword filtering and sentence segmentation. Chemical element symbols and formulas are explicitly preserved to maintain their semantic relevance for materials-related embeddings. The cleaned, tokenised abstracts form the input for training the Word2Vec model. Transformer-based embeddings operate directly as pre-trained encoders and do not require this corpus-level preprocessing.

\subsection{Construction of text-derived composition embeddings}

Each composition in the experimental materials libraries is represented as a vector in a latent ``text space'', and all downstream analysis (similarity to concepts and Pareto filtering) is performed on these vectors. The models differ only in how this composition embedding is constructed; the similarity and Pareto procedures are identical across all methods.

\subsubsection{Word2Vec baseline (W2V)}

As a lightweight baseline, we train word embeddings using the skip-gram architecture implemented in the \texttt{VecGenerator} module of \texttt{MatNexus}, based on the Word2Vec model~\cite{Mikolov2013,mikolov2013efficient,Goldberg2014}. Compared with TF-IDF or large transformer models, Word2Vec provides compact, dense vector representations at comparatively low computational cost, making it suitable for focused scientific corpora~\cite{Rogers2020842}. Training is performed with 200-dimensional vectors, a window size of 5, hierarchical softmax and a minimum word frequency of 1. All available CPU threads are used for parallelisation.

Element symbols (for example, ``Ag'', ``Pd'', ``Pt'') are treated as ordinary tokens in the corpus. For a given binary or multicomponent composition $c$, we first extract the normalised atomic fractions of all present elements from the materials library (for example, Ag\textsubscript{0.5}Pd\textsubscript{0.5}). The corresponding composition vector $\mathbf{v}_\text{W2V}(c)$ is computed as a concentration-weighted linear combination of the element embeddings,
\[
\mathbf{v}_\text{W2V}(c) \;=\; \sum_{i} x_i\,\mathbf{w}_i,
\]
where $x_i$ is the atomic fraction of element $i$ and $\mathbf{w}_i$ is its Word2Vec embedding. The resulting vector is $\ell_2$-normalised prior to similarity calculation. An example of the composition embeddings distribution is shown in Fig.~\ref{S-fig:tsne_cosine_similarity_distribution} a).

\subsubsection{Element-wise transformer embeddings (MatSciBERT and Qwen)}

To probe the benefit of contextual transformer embeddings while retaining the simple linear-combination scheme, we construct element-wise material vectors using two models:

\begin{itemize}
  \item MatSciBERT~\cite{Gupta2022}: a domain-specific transformer model applied to short element sentences.
  \item Qwen~\cite{qwen}: a Qwen-based embedding model accessed via the Blablador service~\cite{Strube:1038640}, using the same element-sentence prompts.
\end{itemize}

For each chemical element $E$ present in the periodic table, we construct a short natural-language prompt of the form ``\texttt{E is a chemical element.}'' and encode this sentence once with the corresponding embedding model. This yields an element embedding $\mathbf{e}_E$ for each element (MatSciBERT and Qwen separately). Rows in the composition data of the experimental materials library tables contain element-wise atomic fractions; these are clipped to non-negative values and renormalised to sum to one for each composition.

For a given composition $c$, such as Ag\textsubscript{0.5}Pd\textsubscript{0.5}, the element-wise transformer vector is defined as the concentration-weighted sum of the element vectors,
\[
\mathbf{v}_\text{elem}(c) \;=\; \sum_{i} x_i\,\mathbf{e}_{E_i},
\]
followed by $\ell_2$ normalisation. This procedure is analogous to the Word2Vec baseline but replaces the distributional word vectors with transformer-derived element embeddings. We refer to the resulting models as MatSciBERT and Qwen throughout the manuscript. Both transformer models are used strictly in inference mode; no additional fine-tuning is performed on our electrocatalysis corpus.

\subsubsection{Composition-prompt transformer embeddings (MatSciBERT\_Full and Qwen\_Full)}

In addition to the element-wise representations, we consider models that operate directly on a textual description of the full composition. For each composition, we construct a prompt of the form
\[
\texttt{material composition: } \mathrm{Ag}=0.50,\, \mathrm{Pd}=0.50,
\]
optionally augmented with simple process tags (for example, co-sputtering, annealing temperature) when available. This prompt encodes both the identity of the elements and their relative fractions as a single piece of text.

Two composition-prompt models are evaluated:

\begin{itemize}
  \item MatSciBERT\_Full: the MatSciBERT model applied directly to the composition prompt.
  \item Qwen\_Full: the Qwen-based Blablador embedding model applied to the same composition prompt.
\end{itemize}

For each row in the library table, the corresponding composition prompt is tokenized and passed through the embedding model. We use mean pooling over the last hidden layer to obtain a single fixed-length vector per composition (MatSciBERT\_Full), or the embedding returned by the Blablador endpoint (Qwen\_Full). All composition vectors are $\ell_2$-normalised. In contrast to the element-wise schemes, these models are free to encode higher-order interactions between elements and their stoichiometric ratios directly in text space.

\subsection{Concept similarity scores}

All models share the same concept layer and downstream similarity calculation. We select two simple, physically motivated descriptors that frequently appear in discussions of electrocatalytic materials and related thin films: \texttt{conductivity} and \texttt{dielectric}~\cite{Foppa20211016}. 

For the Word2Vec model, we use the individual tokens ``conductivity'' and ``dielectric'' and extract their corresponding word embeddings. An example of the resulting similarity-score distribution in the two-concept space is shown in Fig.~\ref{S-fig:tsne_cosine_similarity_distribution} b). For the transformer-based models, we encode short natural-language phrases, such as ``\texttt{conductivity material property}'' and ``\texttt{dielectric material property}'', to obtain concept vectors. In all cases, the concept vectors are $\ell_2$-normalised.

Given a normalised composition vector $\mathbf{v}(c)$ (from W2V, MatSciBERT, Qwen, MatSciBERT\_Full or Qwen\_Full) and a normalised concept vector $\mathbf{v}_k$ (for $k \in \{\text{conductivity}, \text{dielectric}\}$), we compute cosine similarity,
\[
S_k \;=\; \mathbf{v}(c) \cdot \mathbf{v}_k.
\]
Thus, each composition is mapped to a point $(S_\mathrm{dielectric}, S_\mathrm{conductivity})$ in a two-dimensional descriptor space derived purely from text.

\subsection{Pareto-front multi-objective filtering}

Multi-objective optimization is performed via Pareto front analysis in the two-dimensional descriptor space spanned by similarity to the dielectric- and conductivity-related concepts. A composition is considered Pareto-optimal if no other composition achieves equal or better performance across all objectives~\cite{pareto1896cours2v}.

Because the two descriptors correspond to complementary electronic regimes that can both be desirable in electrocatalysts, we do not assign one concept as universally favourable. Instead, we construct two Pareto fronts with opposite objective directions:

\begin{enumerate}
  \item Maximize similarity to \texttt{conductivity} while simultaneously minimizing similarity to \texttt{dielectric};
  \item Maximize similarity to \texttt{dielectric} while simultaneously minimizing similarity to \texttt{conductivity}.
\end{enumerate}

For each embedding model and each material system, non-dominated compositions are identified separately in the corresponding objective space. The final candidate set used for experimental comparison is the union of the two fronts, ensuring a symmetric treatment of the two concepts. The same dual-front construction is applied uniformly to all embedding models (W2V, MatSciBERT, Qwen, MatSciBERT\_Full and Qwen\_Full) and to all alloy and oxide libraries considered in this study.

\section{Results}
\label{sec:results}

\begin{table*}[htbp]
\centering
\caption{Performance of W2V, MatSciBERT, and Qwen across all material systems and overpotentials. Here, count/O and count/P denote the number of compositions in the original dataset and in the Pareto-selected subset, respectively; best/O and best/P are the best current densities in each set (mA\,cm$^{-2}$); error (\%) is the relative deviation between best/P and best/O; and fraction (\%) is the percentage of original candidates retained in the Pareto subset.}
\label{tab:perf-base-models}
\begin{tabular}{l r l r r r r r r}
\hline
Material system & $\eta$ (mV) & Method & count/O & count/P & best/O & best/P & error (\%) & fraction (\%) \\
\hline
\multirow{3}{*}{Ag-Au-Pd-Pt-Rh} & \multirow{3}{*}{-300} 
& MatSciBERT & 327 & 154 & -1.1 & -1.1 & 0.0 & 47.1 \\
& & Qwen  & 327 & 93  & -1.1 & -1.1 & 0.0 & 28.4 \\
& & W2V   & 327 & 9   & -1.1 & -1.1 & 3.8 & 2.8  \\
\hline
\multirow{3}{*}{Ag-Au-Pd-Pt-Ru} & \multirow{3}{*}{-300} 
& MatSciBERT & 335 & 88  & -1.5 & -1.5 & 0.1 & 26.3 \\
& & Qwen  & 335 & 109 & -1.5 & -1.5 & 2.8 & 32.5 \\
& & W2V   & 335 & 20  & -1.5 & -1.5 & 2.8 & 6.0  \\
\hline
\multirow{3}{*}{Ag-Pd-Pt} & \multirow{3}{*}{850} 
& MatSciBERT & 341 & 45 & -0.6 & -0.5 & 15.7 & 13.2 \\
& & Qwen  & 341 & 71 & -0.6 & -0.5 & 11.1 & 20.8 \\
& & W2V   & 341 & 15 & -0.6 & -0.4 & 23.4 & 4.4  \\
\hline
\multirow{3}{*}{Ag-Pd-Pt-Ru} & \multirow{3}{*}{850} 
& MatSciBERT & 341 & 49 & -0.4 & -0.4 & 0.0 & 14.4 \\
& & Qwen  & 341 & 86 & -0.4 & -0.4 & 1.0 & 25.2 \\
& & W2V   & 341 & 13 & -0.4 & -0.4 & 0.0 & 3.8  \\
\hline
\multirow{3}{*}{Ag-Pd-Ru} & \multirow{3}{*}{850} 
& MatSciBERT & 342 & 50 & -0.7 & -0.7 & 0.0 & 14.6 \\
& & Qwen  & 342 & 63 & -0.7 & -0.7 & 2.9 & 18.4 \\
& & W2V   & 342 & 14 & -0.7 & -0.7 & 0.0 & 4.1  \\
\hline
\multirow{3}{*}{Co-Ni-La} & \multirow{3}{*}{1800} 
& MatSciBERT & 342 & 232 & 2.2 & 2.2 & 0.0  & 67.8 \\
& & Qwen  & 342 & 289 & 2.2 & 1.4 & 38.5 & 84.5 \\
& & W2V   & 342 & 66  & 2.2 & 2.2 & 0.0  & 19.3 \\
\hline
\multirow{3}{*}{Cr-Mn-Fe-Co-Ni-O-300C-Per1} & \multirow{3}{*}{1700}
& MatSciBERT & 342 & 303 & 1.9 & 1.8 & 2.2 & 88.6 \\
& & Qwen  & 342 & 66  & 1.9 & 1.8 & 3.6 & 19.3 \\
& & W2V   & 342 & 90  & 1.9 & 1.8 & 2.2 & 26.3 \\
\hline
\multirow{3}{*}{Cr-Mn-Fe-Co-Ni-O-500C-Per1} & \multirow{3}{*}{1700}
& MatSciBERT & 342 & 306 & 2.3 & 2.3 & 0.0 & 89.5 \\
& & Qwen  & 342 & 68  & 2.3 & 2.2 & 4.2 & 19.9 \\
& & W2V   & 342 & 81  & 2.3 & 2.2 & 4.4 & 23.7 \\
\hline
\multirow{3}{*}{Cr-Mn-Fe-Co-Ni-O-500C-Per2} & \multirow{3}{*}{1700}
& MatSciBERT & 342 & 297 & 2.7 & 2.7 & 0.0 & 86.8 \\
& & Qwen  & 342 & 96  & 2.7 & 2.7 & 1.3 & 28.1 \\
& & W2V   & 342 & 71  & 2.7 & 2.7 & 0.9 & 20.8 \\
\hline
\multirow{3}{*}{Cr-Mn-Fe-Co-Ni-O-500C-Per3} & \multirow{3}{*}{1700}
& MatSciBERT & 342 & 243 & 2.2 & 2.2 & 0.0 & 71.1 \\
& & Qwen  & 342 & 104 & 2.2 & 2.1 & 4.1 & 30.4 \\
& & W2V   & 342 & 135 & 2.2 & 2.1 & 2.2 & 39.5 \\
\hline
\multirow{3}{*}{Cr-Mn-Fe-Co-Ni-O-700C-Per1} & \multirow{3}{*}{1700}
& MatSciBERT & 342 & 291 & 1.5 & 1.5 & 0.0 & 85.1 \\
& & Qwen  & 342 & 62  & 1.5 & 1.5 & 0.0 & 18.1 \\
& & W2V   & 342 & 81  & 1.5 & 1.4 & 5.7 & 23.7 \\
\hline
\multirow{3}{*}{Mg-Co-Ni-Cu-O-Per1} & \multirow{3}{*}{1700}
& MatSciBERT & 342 & 47 & 4.9 & 4.9 & 0.0 & 13.7 \\
& & Qwen  & 342 & 51 & 4.9 & 4.9 & 0.0 & 14.9 \\
& & W2V   & 342 & 49 & 4.9 & 4.9 & 0.0 & 14.3 \\
\hline
\multirow{3}{*}{Mg-Co-Ni-Cu-O-Per3} & \multirow{3}{*}{1700}
& MatSciBERT & 342 & 85 & 4.3 & 4.3 & 0.0 & 24.9 \\
& & Qwen  & 342 & 67 & 4.3 & 4.3 & 0.0 & 19.6 \\
& & W2V   & 342 & 97 & 4.3 & 4.3 & 0.0 & 28.4 \\
\hline
\multirow{3}{*}{Ni-Pd-Pt-Ru} & \multirow{3}{*}{1700}
& MatSciBERT & 4026 & 465 & 6.9 & 1.3 & 80.9 & 11.5 \\
& & Qwen  & 4026 & 846 & 6.9 & 6.9 & 0.0  & 21.0 \\
& & W2V   & 4026 & 391 & 6.9 & 6.9 & 0.0  & 9.7  \\
\hline
\end{tabular}
\end{table*}

\begin{table*}[htbp]
\centering
\caption{Performance of MatSciBERT\_Full and Qwen\_Full across all material systems and overpotentials. Here, count/O and count/P denote the number of compositions in the original dataset and in the Pareto-selected subset, respectively; best/O and best/P are the best current densities in each set (mA\,cm$^{-2}$); error (\%) is the relative deviation between best/P and best/O; and fraction (\%) is the percentage of original candidates retained in the Pareto subset.}
\label{tab:perf-full-models}
\begin{tabular}{l r l r r r r r r}
\hline
Material system & $\eta$ (mV) & Method & count/O & count/P & best/O & best/P & error (\%) & fraction (\%) \\
\hline
\multirow{2}{*}{Ag-Au-Pd-Pt-Rh} & \multirow{2}{*}{-300}
& MatSciBERT\_Full & 327 & 37 & -1.1 & -1.1 & 1.3 & 11.3 \\
& & Qwen\_Full    & 327 & 55 & -1.1 & -1.1 & 4.8 & 16.8 \\
\hline
\multirow{2}{*}{Ag-Au-Pd-Pt-Ru} & \multirow{2}{*}{-300}
& MatSciBERT\_Full & 335 & 41 & -1.5 & -1.5 & 2.4 & 12.2 \\
& & Qwen\_Full    & 335 & 74 & -1.5 & -1.5 & 0.4 & 22.1 \\
\hline
\multirow{2}{*}{Ag-Pd-Pt} & \multirow{2}{*}{850}
& MatSciBERT\_Full & 341 & 46 & -0.6 & -0.6 & 0.0  & 13.5 \\
& & Qwen\_Full    & 341 & 61 & -0.6 & -0.5 & 13.9 & 17.9 \\
\hline
\multirow{2}{*}{Ag-Pd-Pt-Ru} & \multirow{2}{*}{850}
& MatSciBERT\_Full & 341 & 50 & -0.4 & -0.3 & 4.7 & 14.7 \\
& & Qwen\_Full    & 341 & 81 & -0.4 & -0.4 & 0.0 & 23.8 \\
\hline
\multirow{2}{*}{Ag-Pd-Ru} & \multirow{2}{*}{850}
& MatSciBERT\_Full & 342 & 60 & -0.7 & -0.7 & 0.0 & 17.5 \\
& & Qwen\_Full    & 342 & 65 & -0.7 & -0.7 & 0.0 & 19.0 \\
\hline
\multirow{2}{*}{Co-Ni-La} & \multirow{2}{*}{1800}
& MatSciBERT\_Full & 342 & 50 & 2.2 & 2.2 & 0.0 & 14.6 \\
& & Qwen\_Full    & 342 & 71 & 2.2 & 2.2 & 2.3 & 20.8 \\
\hline
\multirow{2}{*}{Cr-Mn-Fe-Co-Ni-O-300C-Per1} & \multirow{2}{*}{1700}
& MatSciBERT\_Full & 342 & 39 & 1.9 & 1.8 & 2.2 & 11.4 \\
& & Qwen\_Full    & 342 & 51 & 1.9 & 1.7 & 6.6 & 14.9 \\
\hline
\multirow{2}{*}{Cr-Mn-Fe-Co-Ni-O-500C-Per1} & \multirow{2}{*}{1700}
& MatSciBERT\_Full & 342 & 38 & 2.3 & 2.3 & 0.0 & 11.1 \\
& & Qwen\_Full    & 342 & 38 & 2.3 & 2.2 & 4.2 & 11.1 \\
\hline
\multirow{2}{*}{Cr-Mn-Fe-Co-Ni-O-500C-Per2} & \multirow{2}{*}{1700}
& MatSciBERT\_Full & 342 & 33 & 2.7 & 2.6 & 3.5 & 9.6 \\
& & Qwen\_Full    & 342 & 46 & 2.7 & 2.6 & 3.4 & 13.5 \\
\hline
\multirow{2}{*}{Cr-Mn-Fe-Co-Ni-O-500C-Per3} & \multirow{2}{*}{1700}
& MatSciBERT\_Full & 342 & 52 & 2.2 & 2.1 & 3.8 & 15.2 \\
& & Qwen\_Full    & 342 & 37 & 2.2 & 2.2 & 0.0 & 10.8 \\
\hline
\multirow{2}{*}{Cr-Mn-Fe-Co-Ni-O-700C-Per1} & \multirow{2}{*}{1700}
& MatSciBERT\_Full & 342 & 41 & 1.5 & 1.4 & 3.3 & 12.0 \\
& & Qwen\_Full    & 342 & 40 & 1.5 & 1.3 & 14.5 & 11.7 \\
\hline
\multirow{2}{*}{Mg-Co-Ni-Cu-O-Per1} & \multirow{2}{*}{1700}
& MatSciBERT\_Full & 342 & 46 & 4.9 & 4.5 & 8.7  & 13.5 \\
& & Qwen\_Full    & 342 & 41 & 4.9 & 4.9 & 0.0 & 12.0 \\
\hline
\multirow{2}{*}{Mg-Co-Ni-Cu-O-Per3} & \multirow{2}{*}{1700}
& MatSciBERT\_Full & 342 & 41 & 4.3 & 3.2 & 24.6 & 12.0 \\
& & Qwen\_Full    & 342 & 47 & 4.3 & 3.2 & 25.0 & 13.7 \\
\hline
\multirow{2}{*}{Ni-Pd-Pt-Ru} & \multirow{2}{*}{1700}
& MatSciBERT\_Full & 4026 & 105 & 6.9 & 6.3 & 9.3 & 2.6 \\
& & Qwen\_Full    & 4026 & 123 & 6.9 & 6.9 & 0.1 & 3.1 \\
\hline
\end{tabular}
\end{table*}

\begin{figure*}[htbp]
  \centering
  \includegraphics[width=\textwidth]{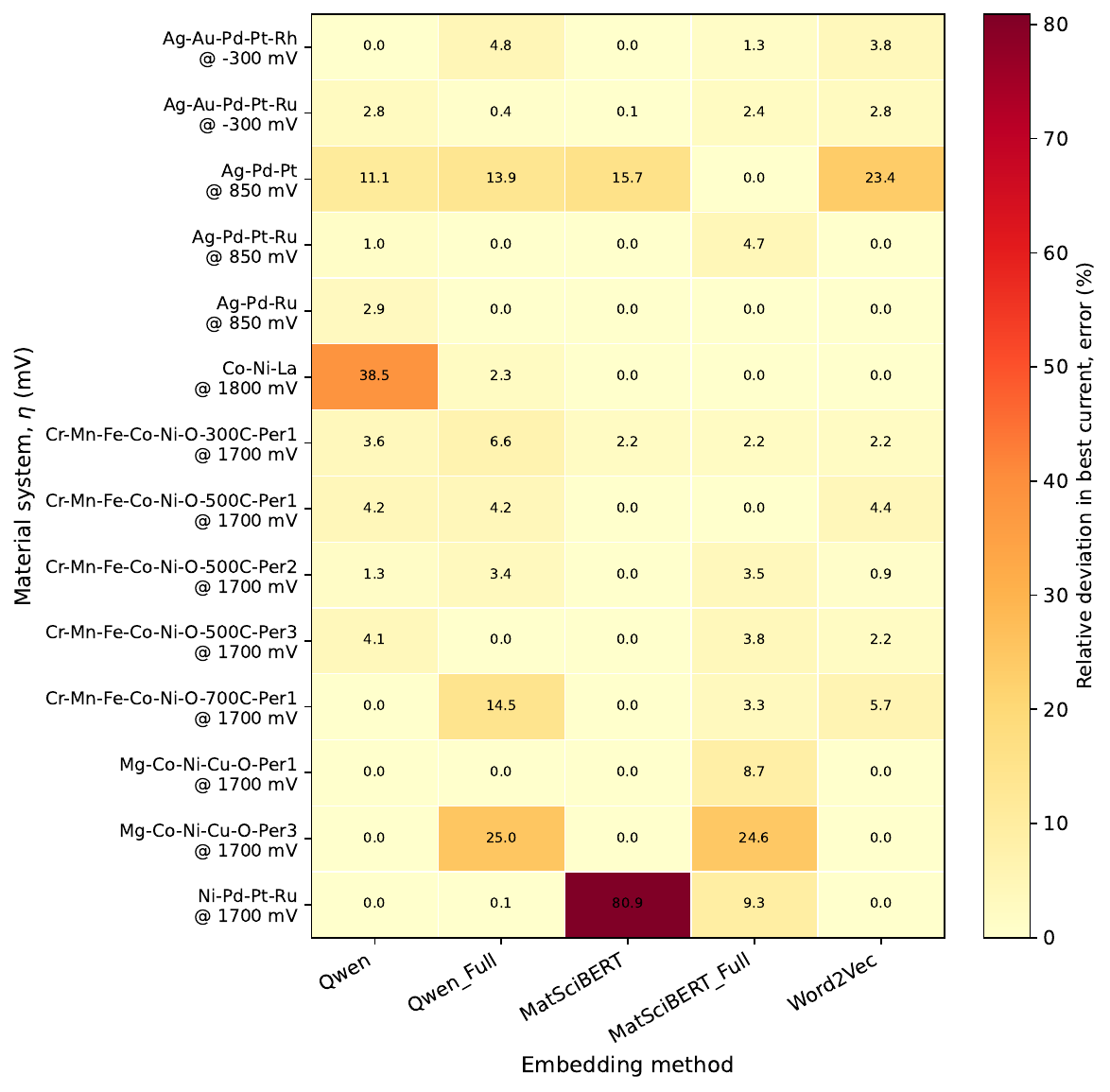}
    \caption{
      Heatmap of the relative error in the current density of the best-performing
      composition between the Pareto-selected subset and the all compositions of the respective materials
      library. Colour indicates the percentage deviation, with lighter shades for
      smaller errors and darker shades for larger errors.}
  \label{fig:pareto_error_heatmap}
\end{figure*}

\begin{figure*}[htbp]
  \centering
  \includegraphics[width=\textwidth]{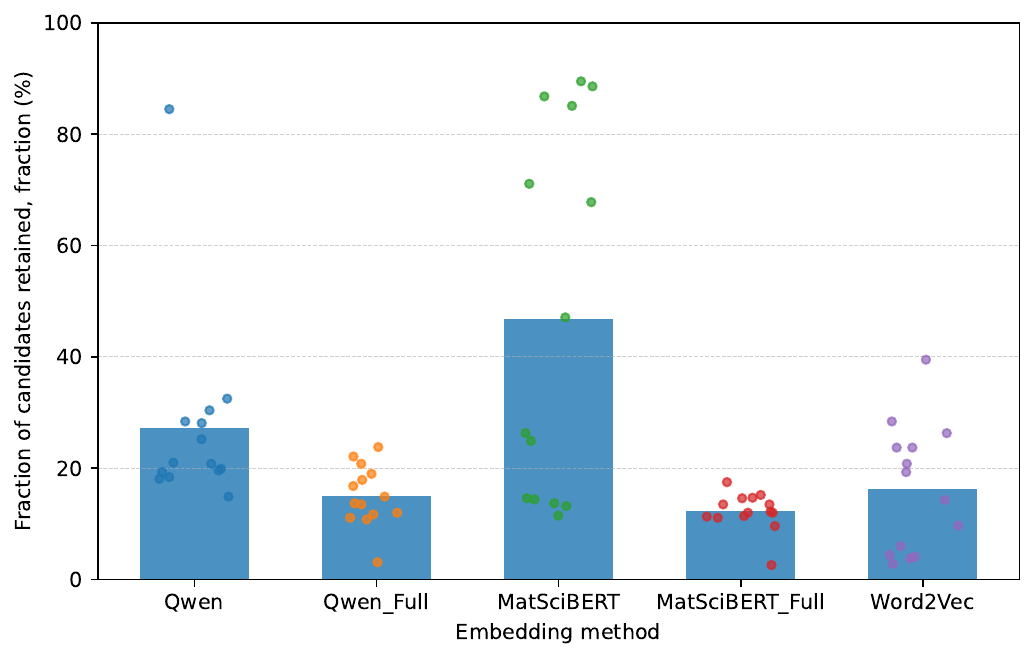}
\caption{Fraction of compositions retained in the Pareto-selected subset for each embedding method. Bars show the mean retained fraction across all material systems. The overlaid dots show the retained fraction for each individual material system; they are slightly shifted left/right only to prevent points from overlapping (the shift has no numerical meaning).}
  \label{fig:fraction_retained_all_methods}
\end{figure*}

\begin{figure*}[htbp]
  \centering
  \includegraphics[width=\textwidth]{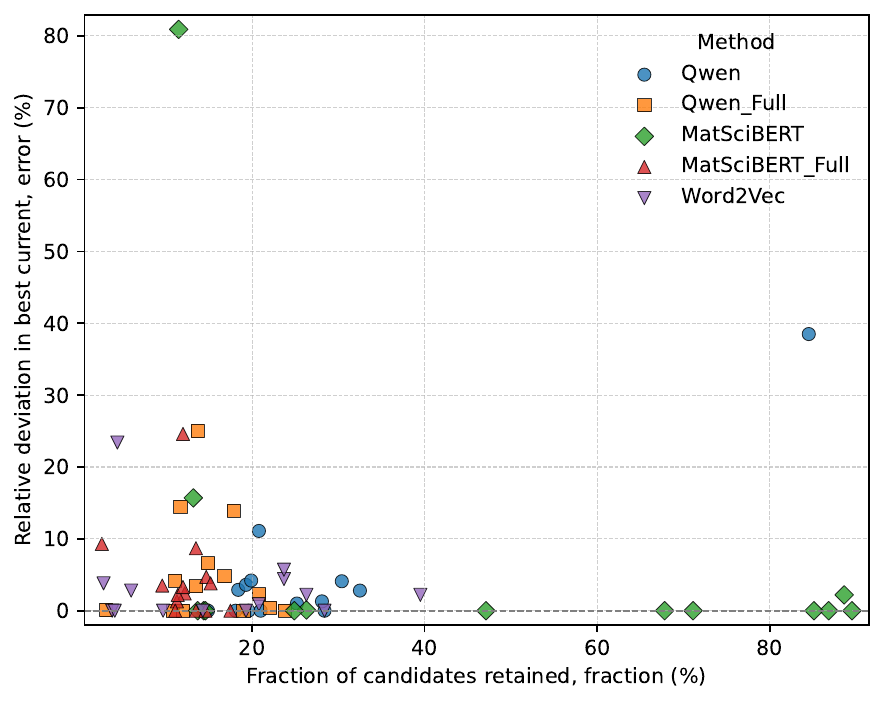}
  \caption{
    Trade-off between the fraction of candidates retained and the relative error in the current density of the best-performing composition for all material systems and embedding
    methods. }
  \label{fig:error_vs_fraction_all_methods}
\end{figure*}

We evaluate five text-derived embedding models on 15 combinatorial materials libraries
covering HER (Ag-Au-Pd-Pt-Ru, Ag-Au-Pd-Pt-Rh), ORR
(Ag-Pd-Pt, Ag-Pd-Pt-Ru, Ag-Pd-Ru) and OER systems. The OER set
includes one metallic library (Ni-Pd-Pt-Ru) and several complex
oxides with different preparation conditions. For each
(system, overpotential, method) triple we report two quantities:
(i) the retained fraction $f_{\mathrm{ret}}$ (\%), defined as the percentage of
original candidates kept in the Pareto-selected subset; and
(ii) the performance deviation $error$ (\%), defined as the relative deviation
between the best current density in the Pareto subset (best/P) and the best
current density in the original library (best/O), as reported in
Tables~\ref{tab:perf-base-models} and~\ref{tab:perf-full-models}.
Thus, $error=0$ indicates that the Pareto subset contains a composition that
matches the original best performer (best/P = best/O)
(Tables~\ref{tab:perf-base-models} and~\ref{tab:perf-full-models},
Figs.~\ref{fig:pareto_error_heatmap}--\ref{fig:error_vs_fraction_all_methods}).
Changes in the spread of compositions are summarised by the IQR
metrics (Table~\ref{S-tab:iqr-narrowing-base-models} and Table~\ref{S-tab:iqr-narrowing-full-models}), and a representative spatial example for the
Ag-Pd-Pt ORR library is shown in
Fig.~\ref{S-fig:AgPdPt_pareto_comparison}.

\subsection{Element-wise embeddings across reaction types}

The element-wise models (W2V, MatSciBERT, Qwen) share the same
linear combination of element vectors but differ in the underlying
text encoder. Across all systems, the Pareto subsets produced by these
methods retain between a few percent and almost the full library
(Table~\ref{tab:perf-base-models} and Table~\ref{S-tab:iqr-narrowing-base-models}, Figs.~\ref{fig:pareto_error_heatmap}
and~\ref{fig:fraction_retained_all_methods}).

For the HER libraries (Ag-Au-Pd-Pt-Ru and Ag-Au-Pd-Pt-Rh),
all three element-wise models keep the most negative current densities
essentially unchanged: $error \leq 4\,\%$
for all methods, with several cases of $error=0$
(Table~\ref{tab:perf-base-models},
Fig.~\ref{fig:pareto_error_heatmap}). The retained fraction spans from
$f_{\mathrm{ret}}\sim 3$--$6\,\%$ for W2V,
through $f_{\mathrm{ret}}\sim 25$--$33\,\%$ for Qwen, up to
$f_{\mathrm{ret}}\sim 26$--$47\,\%$ for MatSciBERT (Table~\ref{tab:perf-base-models},
Fig.~\ref{fig:fraction_retained_all_methods}).

For the ORR libraries (Ag-Pd-Pt, Ag-Pd-Pt-Ru, Ag-Pd-Ru) the same
pattern holds. The performance deviation remains below
$error\sim 25\,\%$ for all element-wise models, and below
$error\sim 12\,\%$ in most cases (Table~\ref{tab:perf-base-models},
Fig.~\ref{fig:pareto_error_heatmap}). W2V selects
$f_{\mathrm{ret}}\sim 4$--$5\,\%$ of compositions, Qwen
$f_{\mathrm{ret}}\sim 18$--$25\,\%$, and MatSciBERT
$f_{\mathrm{ret}}\sim 13$--$15\,\%$ (Table~\ref{tab:perf-base-models},
Fig.~\ref{fig:fraction_retained_all_methods}). The IQR statistics for
these noble-metal systems show mixed behavior: depending on the
library and method, the Pareto subset can either narrow or broaden the
composition distribution (Table~\ref{S-tab:iqr-narrowing-base-models}).

For the oxide OER libraries, MatSciBERT tends to keep a large retained fraction
(typically $f_{\mathrm{ret}}=70$--$90\,\%$ for the
Cr-Mn-Fe-Co-Ni-O systems, and $f_{\mathrm{ret}}=14$--$25\,\%$ for
Mg-Co-Ni-Cu-O), with $error$ in the best current
density mostly below $5\,\%$ (Table~\ref{tab:perf-base-models},
Figs.~\ref{fig:pareto_error_heatmap} and~\ref{fig:error_vs_fraction_all_methods}).
Qwen and W2V select smaller subsets
($f_{\mathrm{ret}}\sim 18$--$40\,\%$ for Qwen and
$f_{\mathrm{ret}}\sim 20$--$40\,\%$ for W2V in the oxides), with
$error$ again generally below $5\,\%$ (Table~\ref{tab:perf-base-models}). The metallic
Ni-Pd-Pt-Ru library is an outlier: MatSciBERT retains
$f_{\mathrm{ret}}\sim 12\,\%$ of candidates but yields a large performance deviation
($error\sim 81\,\%$), whereas Qwen and W2V keep
$f_{\mathrm{ret}}\sim 21\,\%$ and $f_{\mathrm{ret}}\sim 10\,\%$ of candidates,
respectively, with $error\approx 0$ (Table~\ref{tab:perf-base-models},
Fig.~\ref{fig:pareto_error_heatmap}). The IQR results for
the OER systems show that, for many element-wise runs, the Pareto
subsets span a composition range similar to or broader than the
original libraries (often negative IQR narrowing in Table~\ref{S-tab:iqr-narrowing-base-models}).

\subsection{Composition-prompt embeddings across reaction types}

The composition-prompt models (MatSciBERT\_Full, Qwen\_Full) operate
directly on formatted composition strings but follow the same Pareto
selection procedure. Across all material systems, both full models typically retain
$f_{\mathrm{ret}}\sim 10$--$20\,\%$ of candidates, with Qwen\_Full
often keeping slightly more than MatSciBERT\_Full
(Table~\ref{tab:perf-full-models} and Table~\ref{S-tab:iqr-narrowing-full-models},
Fig.~\ref{fig:fraction_retained_all_methods}).

In the HER libraries, both full models achieve small performance deviations,
typically $error \lesssim 5\,\%$, and in several systems $error=0$
(Table~\ref{tab:perf-full-models},
Fig.~\ref{fig:pareto_error_heatmap}). The corresponding retained fractions
lie at $f_{\mathrm{ret}}\sim 11$--$22\,\%$, intermediate between W2V and
the element-wise MatSciBERT approaches (Table~\ref{tab:perf-base-models}
and Table~\ref{tab:perf-full-models}).

For the ORR systems, MatSciBERT\_Full and Qwen\_Full again preserve the
top performance with modest deviation ($error\sim 5$--$15\,\%$) while
retaining $f_{\mathrm{ret}}\sim 13$--$24\,\%$ of candidates
(Table~\ref{tab:perf-full-models}). In the Ag-Pd-Pt library,
MatSciBERT\_Full yields $error=0$ at $f_{\mathrm{ret}}\sim 14\,\%$, whereas
Qwen\_Full shows $error\sim 14\,\%$ at $f_{\mathrm{ret}}\sim 18\,\%$
(Table~\ref{tab:perf-full-models},
Fig.~\ref{fig:error_vs_fraction_all_methods}).

For the OER oxides, both composition-prompt models generally keep the
deviation below $error\sim 10\,\%$, with a larger deviation
($error\sim 25\,\%$) in the Mg-Co-Ni-Cu-O-Per3 library for both
MatSciBERT\_Full and Qwen\_Full (Table~\ref{tab:perf-full-models}).
Retained fractions are $f_{\mathrm{ret}}\sim 10$--$15\,\%$, comparable to
HER and ORR. In the metallic Ni-Pd-Pt-Ru library, MatSciBERT\_Full retains
$f_{\mathrm{ret}}\sim 3\,\%$ with $error\sim 9\,\%$, whereas Qwen\_Full
retains a similar fraction with essentially zero deviation ($error\approx 0$)
(Table~\ref{tab:perf-full-models},
Figs.~\ref{fig:pareto_error_heatmap} and~\ref{fig:error_vs_fraction_all_methods}).
The IQR shifts for the full models remain closer to zero than for the
element-wise models, indicating moderate redistribution of compositions
rather than systematic narrowing of composition ranges
(Table~\ref{S-tab:iqr-narrowing-full-models}).

\subsection{Comparison between element-wise and full embeddings}

Comparing models that share the same backbone allows the assessment of the impact of encoding the compositions as composition-weighted individual elements versus full composition strings.
For the MatSciBERT pair, MatSciBERT usually retains a larger retained fraction than MatSciBERT\_Full, especially in the OER oxides (e.g. $f_{\mathrm{ret}}=70$--$90\,\%$ vs.\ $f_{\mathrm{ret}}\sim 10$--$15\,\%$ in
the Cr-Mn-Fe-Co-Ni-O libraries; Table~\ref{tab:perf-base-models} and Table~\ref{tab:perf-full-models},
Fig.~\ref{fig:fraction_retained_all_methods}). The performance deviations
are small for both, except for two cases:
Ag-Pd-Pt (where MatSciBERT shows $error\sim 16\,\%$ vs.\ $error=0$
for MatSciBERT\_Full), and Ni-Pd-Pt-Ru
($error\sim 81\,\%$ for MatSciBERT vs.\ $error\sim 9\,\%$ for
MatSciBERT\_Full; Table~\ref{tab:perf-base-models} and Table~\ref{tab:perf-full-models},
Figs.~\ref{fig:pareto_error_heatmap} and~\ref{fig:error_vs_fraction_all_methods}).

For the Qwen pair, Qwen typically keeps a larger subset than
Qwen\_Full in the HER and ORR systems, whereas in many oxide OER
systems their retained fractions are closer
(Table~\ref{tab:perf-base-models} and Table~\ref{tab:perf-full-models}, Fig.~\ref{fig:fraction_retained_all_methods}).
The performance deviations are mostly within
$error\sim 5\,\%$ for both, with two larger values: Co-Ni-La
($error\sim 39\,\%$ for Qwen vs.\ $error\sim 2\,\%$ for
Qwen\_Full) and Mg-Co-Ni-Cu-O-Per3
($error\approx 0$ for Qwen vs.\ $error\sim 25\,\%$ for
Qwen\_Full; Table~\ref{tab:perf-base-models} and Table~\ref{tab:perf-full-models},
Figs.~\ref{fig:pareto_error_heatmap} and~\ref{fig:error_vs_fraction_all_methods}).

Relative to the full models, W2V tends to select the smallest subsets
across reaction types (typically $f_{\mathrm{ret}}\sim 3$--$16\,\%$,
with a few oxide systems around $f_{\mathrm{ret}}\sim 25$--$30\,\%$;
Table~\ref{tab:perf-base-models} and Table~\ref{tab:perf-full-models}, Fig.~\ref{fig:fraction_retained_all_methods})
while maintaining performance deviations within
$error\sim 6\,\%$ for all libraries considered (Table~\ref{tab:perf-base-models},
Fig.~\ref{fig:pareto_error_heatmap}). The IQR statistics show that W2V
can either narrow or broaden the composition range depending
on the system, with pronounced narrowing in Ni-Pd-Pt-Ru and several
noble-metal libraries (Table~\ref{S-tab:iqr-narrowing-base-models} and Table~\ref{S-tab:iqr-narrowing-full-models}).

\subsection{Trade-off between candidate reduction and best-current deviation}

The scatter plot of $error$ versus $f_{\mathrm{ret}}$
(Fig.~\ref{fig:error_vs_fraction_all_methods}) summarizes the
trade-offs across all systems and methods. MatSciBERT points cluster
at high retained fractions ($f_{\mathrm{ret}}\gtrsim 40\,\%$) with
performance deviations mostly below $error\sim 10\,\%$, apart from the Ni-Pd-Pt-Ru outlier.
W2V points lie predominantly in the low-retention region
($f_{\mathrm{ret}}\lesssim 20\,\%$) while remaining close to the horizontal axis,
consistent with small deviations. Qwen, MatSciBERT\_Full and Qwen\_Full occupy intermediate
positions in this plane, with retained fractions mostly in the
$f_{\mathrm{ret}}=10$--$30\,\%$ range and deviations usually below
$error\sim 10\,\%$ (Fig.~\ref{fig:error_vs_fraction_all_methods}).
Within this global view, HER, ORR and OER systems are interleaved:
for all three reaction types there exist configurations with small
deviation at low, medium and high retained fractions.

\subsection{Spatial sampling in a representative ORR library}

The Ag-Pd-Pt ORR library illustrates how the Pareto subsets are distributed in real space.
Panel~(a) of Fig.~\ref{S-fig:AgPdPt_pareto_comparison} shows the full $x$-$y$ layout of measured compositions colored by current density.
Panels~(b)-(f) display the corresponding Pareto-selected subsets for MatSciBERT, MatSciBERT\_Full, Qwen, Qwen\_Full and W2V, with the rest of the compositions shown in gray.
In all cases, the Pareto subsets form sparse subsets of the full materials library rather than a tight cluster, with
points chosen across a broad range of $x$ and $y$.
This qualitative pattern is consistent with the IQR results, which show modest narrowing or broadening of the composition distributions in the selected subsets rather than strong contraction to a small region (Table~\ref{S-tab:iqr-narrowing-base-models} and Table~\ref{S-tab:iqr-narrowing-full-models}).

\section{Discussion}
\label{sec:discussion}

We use text-derived composition embeddings, combined with a dual Pareto-front construction, as a filtering tool for combinatorial electrocatalyst materials libraries.
Rather than aiming to predict current densities directly, our method ranks compositions in a two-dimensional descriptor space and selects those that are non-dominated with respect to two concept similarities.

The central question is whether such a filtering scheme based on text mining can discard a large
fraction of candidates while still retaining at least one near-optimal composition in each library. Across HER, ORR and OER systems, our results show that this is often achievable: Pareto subsets
are substantially smaller than the original materials libraries' composition range yet typically
contain a composition whose current density is close to the best-measured value (Figs.~\ref{fig:pareto_error_heatmap}-\ref{fig:error_vs_fraction_all_methods}, Table~\ref{tab:perf-base-models} and Table~\ref{tab:perf-full-models}).

A useful way to interpret these results is to separate the \emph{shared selection pipeline} from the \emph{embedding models} that feed into it.
All methods map each composition to a vector in a latent embedding space, then project this composition-representing vector onto two concept directions (`conductivity' and `dielectric'), and apply the same
two-dimensional Pareto filtering.
For the element-wise variants (W2V, MatSciBERT, Qwen), the composition vector itself is constructed as a linear combination of element embeddings weighted by atomic fraction.
This downstream machinery is deliberately simple: linear mixing, normalization, two scalar similarities, and a Pareto filter.
The transformer-based models differ only in how the element or composition vectors are obtained from text, not in how these vectors are used.
The observation that such a minimal concept layer and linear mixing scheme can still identify subsets that preserve near-best-performing compositions suggests that, for the libraries studied here, `coarse' text-derived representations capture fuzzy correlations that are mappable to electrocatalytic performance.

Within this common framework, comparing the three element-wise models highlights how the choice of encoder modulates the trade-off between filtering strength and robustness.
In both HER and noble-metal ORR libraries, W2V, Qwen and MatSciBERT usually retain at least one
composition with current density close to the best measured value, but the retained fractions differ systematically (Figs.~\ref{fig:pareto_error_heatmap}-\ref{fig:fraction_retained_all_methods}, Table~\ref{tab:perf-base-models}).
W2V tends to return the smallest subsets, Qwen occupies an intermediate regime, and MatSciBERT often retains the most candidate compositions, especially in the oxide OER libraries.
This pattern is consistent with the idea that the three encoders impose broadly similar rankings on the compositions along the conductivity–dielectric axes, but differ in the detailed structure of that ordering.
Small differences in relative similarity then translate into more or fewer points lying on the union of the two Pareto fronts.

The comparison between element-wise and composition-prompt models (MatSciBERT vs.\ MatSciBERT\_Full, Qwen vs.\ Qwen\_Full) shows how embedding the full composition string modifies this balance.
The composition-prompt models are free to encode higher-order interactions between elements and stoichiometric ratios that are inaccessible to a purely linear mixture of independent element vectors.
For some chemical systems, this extra flexibility appears beneficial w.r.t. candidate composition prediction: the full models can match or slightly improve the quality of the best-selected composition at similar or lower retained fractions, with the Ni-Pd-Pt-Ru OER system providing a clear example where MatSciBERT\_Full avoids the large deviation seen for element-wise MatSciBERT (Table~\ref{tab:perf-base-models} and Table~\ref{tab:perf-full-models}).
In other chemical systems, the full models behave similar to their element-wise counterparts, or show modestly larger deviations without a clear gain in filtering.
These mixed outcomes indicate that composition prompts can help when the response depends on
specific multi-element motifs or ratios that are well represented in text, but they are not uniformly advantageous across all reaction types and chemical systems.

Looking across reaction types, our method behaves most consistently in the HER and noble-metal ORR libraries.
Here, the literature is rich in discussions of metallic conductivity, alloying trends and surface chemistry, and these topics are likely to be well captured in the electrocatalysis corpus used for training the embeddings.
It is plausible that similarity to conductivity- and dielectric-related concepts aligns more directly with the underlying physics in these systems, so that even a coarse two-concept descriptor space is sufficient to highlight compositions that are experimentally active.
The oxide OER libraries present a more demanding setting: activity is strongly influenced by oxidation state, local coordination, defects and transport through complex perovskite or spinel structures, which are only partially reflected in element names and simple composition-encoding strings.
In this regime, element-wise MatSciBERT tends to keep a large fraction of the candidates and achieves small deviations, whereas W2V and Qwen more aggressively reduce the library at the cost of slightly larger errors (Table~\ref{tab:perf-base-models}, Fig.~\ref{fig:error_vs_fraction_all_methods}).
The metallic Ni-Pd-Pt-Ru OER library highlights a different failure mode: element-wise MatSciBERT does not contain the best-performing composition, while W2V, Qwen and the full models avoid this bias.
These observations support an interpretation in which the alignment between text-derived concept space and physical response is controlled jointly by the encoder, the corpus and the material class: good performance in one family of systems does not necessarily translate to similarly good behavior elsewhere.

The IQR analysis and the Ag-Pd-Pt spatial plots clarify how the Pareto filters the composition space of materials libraries.
Across most model–system combinations, the elemental interquartile ranges in the Pareto subsets are comparable to those in the full libraries, and in several cases they are slightly broader (Table~\ref{S-tab:iqr-narrowing-base-models} and Table~\ref{S-tab:iqr-narrowing-full-models}).
The spatial distributions for Ag-Pd-Pt (Fig.~\ref{S-fig:AgPdPt_pareto_comparison}) show that the selected points remain spread over the materials library: each model retains compositions along different arcs or clusters, but none collapses onto a single, narrow region.
Gray points emphasize that most measured compositions are removed, yet the surviving candidates still cover multiple distinct parts of the materials library.
This suggests that the method acts mainly as a point-wise filter that favors compositions with extreme values among the candidate compositions, rather than as a tool that sharply contracts the explored composition range.
For exploratory screening, such behavior is attractive: it reduces the number of experiments while maintaining diversity in the retained compositions.

A recurring theme is the competitiveness of the simple W2V baseline.
Despite exhibiting a comparatively small architecture and being trained on a small corpus and using only distributional statistics over element tokens, W2V often matches the transformer-based models in terms of retaining a near-best-performing composition at a given level of filtering, and in
some libraries it offers a slightly better trade-off between error and retained fraction (Figs.~\ref{fig:fraction_retained_all_methods} and~\ref{fig:error_vs_fraction_all_methods}).
Because all models share the same linear mixing and Pareto pipeline, this points to the corpus statistics and the coarse choice of concept directions as the dominant factors for this task, with the additional representational capacity of transformers playing a secondary role.
In practical terms, this supports the use of W2V-derived element embeddings as a reasonable, low-cost baseline for constructing text-derived descriptors of composition space, with transformer-based models providing an optional route in settings where their extra flexibility can be justified.

Our implementation nonetheless contains several deliberate simplifications that shape both its strengths and its limitations.
The descriptor space is restricted to two concept directions and does not explicitly account for other factors known to influence electrocatalyst performance, such as stability, adsorption energy, or detailed structural information beyond the coarse processing tags included in the composition prompts.
The evaluation focuses on the best current density at a single overpotential per system, so it does not probe how well the filter preserves other features of the electrochemical performance landscape.
The embeddings are also influenced by how they are obtained.
For the W2V baseline, the vectors depend directly on the electrocatalysis-focused corpus assembled in our work, as well as on the chosen training hyperparameters.
In contrast, the transformer-based embeddings are taken from pre-trained MatSciBERT and Qwen models, whose underlying training data and objectives are defined externally and only partly documented.
In all cases, we fixed the W2V corpus and the prompt templates used to generate element-wise or composition-prompt embeddings; the questions how a different corpora, possible pre-training mixtures or different prompt formulations might alter the behavior of the filtering process remains to be explored.
Finally, the size of the Pareto subset is an emergent property of the objective space and cannot currently be tuned directly, and the method does not provide uncertainty estimates or formal guaranties on the retained error of the prediction.

Within these boundaries, our present results indicate that text-derived composition embeddings, combined with a simple concept layer and Pareto-front filtering, offer a practical way to reduce the candidate composition space of combinatorial materials libraries for electrochemical applications while retaining high-performing candidates for HER, ORR and OER.
The fact that linear combinations of element embeddings from a Word2Vec model already perform well suggests that relatively modest text-derived representations can be sufficient for this kind of pre-screening, and that composition-prompt transformer embeddings provide an additional but not strictly necessary layer.
In this sense, the approach does not replace detailed mechanistic modeling or supervised learning on experimental data but can complement them by providing an inexpensive, label-free filter that substantially narrows the experimental search space while retaining good-to-best-performing compositions across diverse material systems as high-probability search directions.

\section{Conclusions}

Text-derived composition embeddings combined with a two-concept similarity map and a dual Pareto-front filter provide a practical, label-free way to reduce combinatorial electrocatalyst libraries while retatining high-performing composition candidates across the hydrogen evolution reaction (HER), oxygen reduction reaction (ORR), and oxygen evolution reaction (OER).
Across 15 noble-metal alloy and multicomponent oxide materials libraries, the different embedding choices occupy distinct regions of the trade-off between candidate reduction and robustness: element-wise, MatSciBERT often retains large fractions of the library, Qwen and the composition-prompt variants typically lie in an intermediate regime, and the lightweight Word2Vec baseline frequently achieves the largest reductions while staying close to the best measured performance.
Comparisons between element-wise and composition-prompt transformers show that encoding full composition strings can improve robustness in specific systems but is not uniformly superior.
Our results indicate that screening performance is governed jointly by the embedding model and by how stoichiometry is represented, and that simple text-based embeddings can serve as part of an effective filtering strategy that narrows experimental search spaces while keeping compositional diversity for follow-up study.

\section*{Author Contributions}
\textbf{Lei Zhang:} Conceptualization, Methodology, Software, Validation, Formal analysis, Investigation, Data Curation, Writing - Original Draft, Visualization, Experimentation, Funding acquisition.\\

\textbf{Markus Stricker:} Conceptualization, Resources, Supervision, Writing - Review \& Editing, Funding acquisition.

\begin{acknowledgments}
Both authors acknowledge inspiring discussions with Maribel Acosta (TUM) and Alfred Ludwig (RUB).
We also acknowledge funding by the Deutsche Forschungsgemeinschaft (DFG, German Research Foundation) under CRC~1625, project number~506711657 (subprojects INF, A05). Lei Zhang acknowledges funding by the DFG through TRR~247, project number 388390466 (subproject INF).
\end{acknowledgments}

\section*{Data and Code Availability}
The code required to reproduce the results of this study is publicly available at \url{https://github.com/lab-mids/w2v_transformers_materials_embeddings#}. The full experimental datasets used in this study are publicly available via Zenodo: ORR dataset at \url{https://doi.org/10.5281/zenodo.13992986}, HER dataset at \url{https://doi.org/10.5281/zenodo.14959252}, and OER dataset (Ni--Pd--Pt--Ru) at \url{https://doi.org/10.5281/zenodo.14891704}.
The Cr--Mn--Fe--Co--Ni--O series dataset is available from the corresponding publication at \url{https://doi.org/10.1021/acs.chemmater.2c01455}. Additional material-system files will be released publicly in a subsequent update of the repository/Zenodo record.

\section*{References}

\nocite{*}
\bibliography{literature}

@inproceedings{devlin-etal-2019-bert,
    title = "{BERT}: Pre-training of Deep Bidirectional Transformers for Language Understanding",
    author = "Devlin, Jacob  and
      Chang, Ming-Wei  and
      Lee, Kenton  and
      Toutanova, Kristina",
    editor = "Burstein, Jill  and
      Doran, Christy  and
      Solorio, Thamar",
    booktitle = "Proceedings of the 2019 Conference of the North {A}merican Chapter of the Association for Computational Linguistics: Human Language Technologies, Volume 1 (Long and Short Papers)",
    month = jun,
    year = "2019",
    address = "Minneapolis, Minnesota",
    publisher = "Association for Computational Linguistics",
    url = "https://aclanthology.org/N19-1423/",
    doi = "10.18653/v1/N19-1423",
    pages = "4171--4186"
}

@inproceedings{levy_goldberg_2014_sgns_pmi,
  title     = {Neural Word Embedding as Implicit Matrix Factorization},
  author    = {Levy, Omer and Goldberg, Yoav},
  booktitle = {Advances in Neural Information Processing Systems (NeurIPS)},
  year      = {2014}
}

@article{Ludwig2019Discovery,
  author  = {Ludwig, Alfred},
  title   = {Discovery of new materials using combinatorial synthesis and high-throughput characterization of thin-film materials libraries combined with computational methods},
  journal = {npj Computational Materials},
  year    = {2019},
  volume  = {5},
  pages   = {70},
  doi     = {10.1038/s41524-019-0205-0},
  publisher = {Springer Nature}
}

@ARTICLE{Löffler202126894,
	author = {Löffler, Tobias and Ludwig, Alfred and Rossmeisl, Jan and Schuhmann, Wolfgang},
	title = {What Makes High-Entropy Alloys Exceptional Electrocatalysts?},
	year = {2021},
	journal = {Angewandte Chemie - International Edition},
	volume = {60},
	number = {52},
	pages = {26894 - 26903},
	doi = {10.1002/anie.202109212},
	url = {https://www.scopus.com/inward/record.uri?eid=2-s2.0-85116126034&doi=10.1002%2Fanie.202109212&partnerID=40&md5=2785c4920e36246e43844f4084457d5f},
	publication_stage = {Final},
	note = {All Open Access; Green Accepted Open Access; Green Open Access; Hybrid Gold Open Access}
}

@ARTICLE{Yao20206316,
	author = {Yao, Yonggang and Huang, Zhennan and Li, Tangyuan and Wang, Hang and Liu, Yifan and Stein, H. S. and Mao, Yimin and Gao, Jinlong and Jiao, Miaolun and Dong, Qi and Dai, Jiaiqi and Xie, Pengfei and Xie, Hua and Lacey, Steven David and Takeuchi, Ichiro and Gregoire, John Mathew and Jiang, Rongzhong and Wang, Chao and Taylor, André D. and Shahbazian-Yassar, Reza and Hu, Liangbing},
	title = {High-throughput, combinatorial synthesis of multimetallic nanoclusters},
	year = {2020},
	journal = {Proceedings of the National Academy of Sciences of the United States of America},
	volume = {117},
	number = {12},
	pages = {6316 - 6322},
	doi = {10.1073/pnas.1903721117},
	url = {https://www.scopus.com/inward/record.uri?eid=2-s2.0-85082322354&doi=10.1073%2Fpnas.1903721117&partnerID=40&md5=9a685bed39f9a41d823f5fc1da88d3d2},
	publication_stage = {Final},
	note = {All Open Access; Green Accepted Open Access; Green Open Access; Hybrid Gold Open Access}
}

@ARTICLE{Baeck2002563,
	author = {Baeck, Sung Hyeon and Jaramillo, Thomas Francisco and Brändli, Christof and McFarland, Eric W.},
	title = {Combinatorial Electrochemical Synthesis and Characterization of Tungsten-Based Mixed-Metal Oxides},
	year = {2002},
	journal = {Journal of Combinatorial Chemistry},
	volume = {4},
	number = {6},
	pages = {563 - 568},
	doi = {10.1021/cc020014w},
	url = {https://www.scopus.com/inward/record.uri?eid=2-s2.0-0042370390&doi=10.1021%2Fcc020014w&partnerID=40&md5=c72c68843a420e71cbcca1d6c0daa834},
	publication_stage = {Final}
}

@ARTICLE{Pandit20227583,
	author = {Pandit, Neeraj Kumar and Roy, Diptendu and Mandal, Shyama Charan and Pathak, Biswarup},
	title = {Rational Designing of Bimetallic/Trimetallic Hydrogen Evolution Reaction Catalysts Using Supervised Machine Learning},
	year = {2022},
	journal = {Journal of Physical Chemistry Letters},
	volume = {13},
	number = {32},
	pages = {7583 - 7593},
	doi = {10.1021/acs.jpclett.2c01401},
	url = {https://www.scopus.com/inward/record.uri?eid=2-s2.0-85136659010&doi=10.1021%2Facs.jpclett.2c01401&partnerID=40&md5=adaed69067261aaf8df4c49d4b6960fc},
	publication_stage = {Final}
}

@ARTICLE{Liu2021,
	author = {Liu, Xinghui and Zheng, Lirong and Han, Chenxu and Zong, Hongxiang and Yang, Guang and Lin, Shiru and Kumar, Ashwani and Jadhav, Amol R. and Tran, Ngoc Quang and Hwang, Yosep and Lee, Jinsun and Vasimalla, Suresh and Chen, Zhongfang and Kim, Seong-gon and Lee, Hyoyoung},
	title = {Identifying the Activity Origin of a Cobalt Single-Atom Catalyst for Hydrogen Evolution Using Supervised Learning},
	year = {2021},
	journal = {Advanced Functional Materials},
	volume = {31},
	number = {18},
	pages = {},
	doi = {10.1002/adfm.202100547},
	url = {https://www.scopus.com/inward/record.uri?eid=2-s2.0-85101260373&doi=10.1002%2Fadfm.202100547&partnerID=40&md5=fbbddc4459c101d5c98b00f4d1f25529},
	publication_stage = {Final}
}

@ARTICLE{Mahbub2020,
	author = {Mahbub, Rubayyat and Huang, Kevin J. and Jensen, Zach and Hood, Zachary David and Rupp, Jennifer L.M. and Olivetti, Elsa A.},
	title = {Text mining for processing conditions of solid-state battery electrolyte},
	year = {2020},
	journal = {Electrochemistry Communications},
	volume = {121},
	pages = {},
	doi = {10.1016/j.elecom.2020.106860},
	url = {https://www.scopus.com/inward/record.uri?eid=2-s2.0-85096178542&doi=10.1016%2Fj.elecom.2020.106860&partnerID=40&md5=33fb42abe5a44325c36a993525d7d033},
	publication_stage = {Final},
	note = {All Open Access; Gold Open Access; Green Accepted Open Access; Green Open Access}
}

@ARTICLE{Tshitoyan201995,
	author = {Tshitoyan, Vahe and Dagdelen, John M. and Weston, Leigh and Dunn, Alexander and Rong, Ziqin and Kononova, Olga and Persson, Kristin A. and Ceder, Gerbrand and Jain, Anubhav},
	title = {Unsupervised word embeddings capture latent knowledge from materials science literature},
	year = {2019},
	journal = {Nature},
	volume = {571},
	number = {7763},
	pages = {95 - 98},
	doi = {10.1038/s41586-019-1335-8},
	url = {https://www.scopus.com/inward/record.uri?eid=2-s2.0-85068420677&doi=10.1038%2Fs41586-019-1335-8&partnerID=40&md5=2d3794e12bc366e5468a0ff09607f24e},
	type = {Article},
	publication_stage = {Final},
	source = {Scopus},
	note = {All Open Access; Green Accepted Open Access; Green Open Access}
}

@misc{zhang2025electrocatalyst,
      title={Electrocatalyst discovery through text mining and multi-objective optimization}, 
      author={Lei Zhang and Markus Stricker},
      year={2025},
      eprint={2502.20860},
      archivePrefix={arXiv},
      primaryClass={cond-mat.mtrl-sci},
      url={https://arxiv.org/abs/2502.20860}, 
      doi={10.48550/arXiv.2502.20860}
}

@ARTICLE{Sun20247392,
	author = {Sun, Xiaowen and Araujo, Rafael B. and Dos Santos, Egon Campos and Sang, Yuanhua and Liu, Hong and Yu, Xiaowen},
	title = {Advancing electrocatalytic reactions through mapping key intermediates to active sites via descriptors},
	year = {2024},
	journal = {Chemical Society Reviews},
	volume = {53},
	number = {14},
	pages = {7392 - 7425},
	doi = {10.1039/d3cs01130e},
	url = {https://www.scopus.com/inward/record.uri?eid=2-s2.0-85196483572&doi=10.1039%2Fd3cs01130e&partnerID=40&md5=0eae714a8f608b0cda350ca12ad0da1d},
	publication_stage = {Final}
}

@ARTICLE{Chen201711969,
	author = {Chen, Dawei and Dong, Chungli and Zou, Yuqin and Su, Dong and Huang, Yucheng and Tao, Li and Dou, Shuo and Shen, Shao Hua and Wang, Shuangyin},
	title = {In situ evolution of highly dispersed amorphous CoOx clusters for oxygen evolution reaction},
	year = {2017},
	journal = {Nanoscale},
	volume = {9},
	number = {33},
	pages = {11969 - 11975},
	doi = {10.1039/c7nr04381c},
	url = {https://www.scopus.com/inward/record.uri?eid=2-s2.0-85028439907&doi=10.1039%2Fc7nr04381c&partnerID=40&md5=e35b0dc5fd37a35e20c0a2323513a292},
	publication_stage = {Final},
	note = {All Open Access; Green Accepted Open Access; Green Open Access}
}

@ARTICLE{Qiao201612658,
	author = {Qiao, Mo and Tang, Cheng and He, Guanjie and Qiu, Kaipei and Binions, Russell and Parkin, Ivan P. and Zhang, Qiang and Guo, Zheng Xiao and Titirici, Maria Magdalena},
	title = {Graphene/nitrogen-doped porous carbon sandwiches for the metal-free oxygen reduction reaction: Conductivity: versus active sites},
	year = {2016},
	journal = {Journal of Materials Chemistry A},
	volume = {4},
	number = {32},
	pages = {12658 - 12666},
	doi = {10.1039/c6ta04578b},
	url = {https://www.scopus.com/inward/record.uri?eid=2-s2.0-84981352257&doi=10.1039%2Fc6ta04578b&partnerID=40&md5=309287316d2371caff615c367d2459d7},
	publication_stage = {Final}
}

@ARTICLE{Qu2024,
	author = {Qu, Jiaxing and Xie, Yuxuan Richard and Ciesielski, Kamil M. and Porter, Claire E. and Toberer, Eric S. and Ertekin, Elif},
	title = {Leveraging language representation for materials exploration and discovery},
	year = {2024},
	journal = {npj Computational Materials},
	volume = {10},
	number = {1},
	pages = {},
	doi = {10.1038/s41524-024-01231-8},
	url = {https://www.scopus.com/inward/record.uri?eid=2-s2.0-85188336206&doi=10.1038%2Fs41524-024-01231-8&partnerID=40&md5=cf4807ef49052803005857cda5cb01fd},
	publication_stage = {Final},
	note = {All Open Access; Gold Open Access}
}

@ARTICLE{Lei20241257,
	author = {Lei, Ge and Docherty, Ronan and Cooper, Samuel J.},
	title = {Materials science in the era of large language models: a perspective},
	year = {2024},
	journal = {Digital Discovery},
	volume = {3},
	number = {7},
	pages = {1257 - 1272},
	doi = {10.1039/d4dd00074a},
	url = {https://www.scopus.com/inward/record.uri?eid=2-s2.0-85196014719&doi=10.1039%2Fd4dd00074a&partnerID=40&md5=cd4add740a20ca70b19bed2b204b2609},
	publication_stage = {Final},
	note = {All Open Access; Gold Open Access}
}

@ARTICLE{Park2024601,
	author = {Park, Hyunsoo and Onwuli, Anthony and Butler, Keith T. and Walsh, Aron},
	title = {Mapping inorganic crystal chemical space},
	year = {2024},
	journal = {Faraday Discussions},
	volume = {256},
	pages = {601 – 613},
	doi = {10.1039/d4fd00063c},
	url = {https://www.scopus.com/inward/record.uri?eid=2-s2.0-85194932074&doi=10.1039%2fd4fd00063c&partnerID=40&md5=5a7f0678d02be10ea45bc7e604bff7e3},
	note = {Cited by: 12}
}

@ARTICLE{Liu2024240,
	author = {Liu, Siyu and Wen, Tongqi and Pattamatta, A. S.L.Subrahmanyam and Srolovitz, David J.},
	title = {A prompt-engineered large language model, deep learning workflow for materials classification},
	year = {2024},
	journal = {Materials Today},
	volume = {80},
	pages = {240 - 249},
	doi = {10.1016/j.mattod.2024.08.028},
	url = {https://www.scopus.com/inward/record.uri?eid=2-s2.0-85204408904&doi=10.1016%2Fj.mattod.2024.08.028&partnerID=40&md5=87318db6c7eb3c6b7992164a94e2b3bb},
	publication_stage = {Final}
}

@ARTICLE{Zhang2024,
	author = {Zhang, Lei and Stricker, Markus},
	title = {MATNEXUS: A comprehensive text mining and analysis suite for materials discovery},
	year = {2024},
	journal = {SoftwareX},
	volume = {26},
	doi = {10.1016/j.softx.2024.101654},
	url = {https://www.scopus.com/inward/record.uri?eid=2-s2.0-85185298843&doi=10.1016%2fj.softx.2024.101654&partnerID=40&md5=014a8011ed5503349a7456df4f31c898},
    pages = {101654},
	type = {Article},
	publication_stage = {Final},
	source = {Scopus},

}

@Article{Mikolov2013,
  author       = {Mikolov, Tomas and Chen, Kai and Corrado, Greg and Dean, Jeffrey and Sutskever, L and Zweig, Geoffrey},
  title        = {word2vec},
  journal      = {URL https://code.google.com/p/word2vec},
  year         = {2013},
  volume       = {22},
  pages        = {795},
  howpublished = {https://code.google.com/p/word2vec},
}

@inproceedings{mikolov2013efficient,
  title        = {Efficient Estimation of Word Representations in Vector Space},
  author       = {Mikolov, Tomas and Chen, Kai and Corrado, Greg and Dean, Jeffrey},
  booktitle    = {Proceedings of the International Conference on Learning Representations (ICLR), Workshop Track},
  year         = {2013},
  url          = {https://arxiv.org/abs/1301.3781}
}

@Misc{Goldberg2014,
  author        = {Yoav Goldberg and Omer Levy},
  title         = {word2vec Explained: deriving Mikolov et al.'s negative-sampling word-embedding method},
  howpublished  = {arXiv:1402.3722},
  year          = {2014},
  archiveprefix = {arXiv},
  eprint        = {1402.3722},
  primaryclass  = {cs.CL},
  url           = {https://arxiv.org/abs/1402.3722},
}

@ARTICLE{Rogers2020842,
	author = {Rogers, Anna and Kovaleva, Olga and Rumshisky, Anna},
	title = {A primer in bertology: What we know about how bert works},
	year = {2020},
	journal = {Transactions of the Association for Computational Linguistics},
	volume = {8},
	pages = {842 – 866},
	doi = {10.1162/tacl_a_00349},
	url = {https://www.scopus.com/inward/record.uri?eid=2-s2.0-85098839172&doi=10.1162%2ftacl_a_00349&partnerID=40&md5=c14ea3cd386b168b85cd64948cf50d91},
	type = {Article},
	publication_stage = {Final},
	source = {Scopus},

}

@ARTICLE{Foppa20211016,
	author = {Foppa, Lucas and Ghiringhelli, Luca M. and Girgsdies, Frank and Hashagen, Maike and Kube, Pierre and Hävecker, Michael and Carey, Spencer J. and Tarasov, Andrey and Kraus, Peter and Rosowski, Frank and Schlögl, Robert and Trunschke, Annette and Scheffler, Matthias},
	title = {Materials genes of heterogeneous catalysis from clean experiments and artificial intelligence},
	year = {2021},
	journal = {MRS Bulletin},
	volume = {46},
	number = {11},
	pages = {1016 – 1026},
	doi = {10.1557/s43577-021-00165-6},
	url = {https://www.scopus.com/inward/record.uri?eid=2-s2.0-85114098001&doi=10.1557%2fs43577-021-00165-6&partnerID=40&md5=cec21b8636e60c7d682c196d3f410954},
	type = {Article},
	publication_stage = {Final},
	source = {Scopus},

}

@book{pareto1896cours2v,
  title     = {Cours d’{É}conomie Politique profess{é} {\`a} l’Universit{é} de Lausanne},
  author    = {Pareto, Vilfredo},
  volume    = {I--II},
  year      = {1896--1897},
  publisher = {F. Rouge},
  address   = {Lausanne, Switzerland},
  pages     = {430, 426},

}

@ARTICLE{Gupta2022,
	author = {Gupta, Tanishq and Zaki, Mohd and Krishnan, N. M.Anoop and Mausam, null},
	title = {MatSciBERT: A materials domain language model for text mining and information extraction},
	year = {2022},
	journal = {npj Computational Materials},
	volume = {8},
	number = {1},
	pages = {},
	doi = {10.1038/s41524-022-00784-w},
	url = {https://www.scopus.com/inward/record.uri?eid=2-s2.0-85129234943&doi=10.1038%2Fs41524-022-00784-w&partnerID=40&md5=c6ba4ec295bfd659210fb90794eaf9d6},
	publication_stage = {Final},
	note = {All Open Access; Gold Open Access}
}

@misc{Strube:1038640,
      author       = {Strube, Alexandre},
      title        = {{H}elmholtz {B}lablador and the {LLM} models' ecosystem},
      school       = {Maritime University of Szczecin},
      reportid     = {FZJ-2025-01611},
      year         = {2024},
      month         = {Apr},
      date          = {2025-04-26},
      organization  = {EuroSciPy 2024, Szczecin (Poland), 26
                       Apr 2025 - 30 Apr 2025},
      subtyp        = {Other},
      cin          = {JSC},
      cid          = {I:(DE-Juel1)JSC-20090406},
      pnm          = {5112 - Cross-Domain Algorithms, Tools, Methods Labs (ATMLs)
                      and Research Groups (POF4-511) / Helmholtz AI Consultant
                      Team FB Information (E54.303.11)},
      pid          = {G:(DE-HGF)POF4-5112 / G:(DE-Juel-1)E54.303.11},
      typ          = {PUB:(DE-HGF)6},
      url          = {https://juser.fz-juelich.de/record/1038640},
}

@article{qwen,
  title={Qwen Technical Report},
  author={Jinze Bai and Shuai Bai and Yunfei Chu and Zeyu Cui and Kai Dang and Xiaodong Deng and Yang Fan and Wenbin Ge and Yu Han and Fei Huang and Binyuan Hui and Luo Ji and Mei Li and Junyang Lin and Runji Lin and Dayiheng Liu and Gao Liu and Chengqiang Lu and Keming Lu and Jianxin Ma and Rui Men and Xingzhang Ren and Xuancheng Ren and Chuanqi Tan and Sinan Tan and Jianhong Tu and Peng Wang and Shijie Wang and Wei Wang and Shengguang Wu and Benfeng Xu and Jin Xu and An Yang and Hao Yang and Jian Yang and Shusheng Yang and Yang Yao and Bowen Yu and Hongyi Yuan and Zheng Yuan and Jianwei Zhang and Xingxuan Zhang and Yichang Zhang and Zhenru Zhang and Chang Zhou and Jingren Zhou and Xiaohuan Zhou and Tianhang Zhu},
  journal={arXiv preprint arXiv:2309.16609},
  year={2023}
}
\bibliographystyle{unsrt}

\end{document}